\documentclass[letterpaper,9pt]{extarticle}  
\usepackage{import}
\usepackage{local}
\usepackage[left=.65in,top=.65in,bottom=.65in,right=.65in]{geometry}
\usepackage{authblk}

\title{Vortex ring beams in nonlinear $\mathcal{PT}$-symmetric  systems}
\author[1,*]{Cristian Mejía-Cortés \orcidlink{0000-0002-6444-2700}}
\author[1]{Jesús Muñoz-Muñoz}
\author[2]{Mario I. Molina \orcidlink{0000-0001-9785-793X}}

\affil[1]{Programa de Física, Facultad de Ciencias Básicas, 
Universidad del Atlántico, Puerto Colombia 081007, Colombia}
\affil[2]{Departamento de Física, Facultad de Ciencias, Universidad 
de Chile, Casilla 653, Santiago, Chile}
\affil[*]{Corresponding author: ccmejia@googlemail.com}

\date{}
\begin{document}
\maketitle
\thispagestyle{empty}

\section{Abstract}

\begin{doublespacing}

\noindent
\textbf{\textcolor{Accent}{In this paper, we investigate a two-dimensional photonic array featuring a
circular shape and an alternating gain and loss pattern. Our analysis revolves
around determining the presence and resilience of optical ring modes with varying
vorticity values. This investigation is conducted with respect to both the
array's length and the strength of the non-Hermitian parameter. For larger
values of array's length, we observe a reduction in the stability domain as the
non-Hermitian parameter increases. Interestingly, upon an increasing 
of vorticity of the optical modes full stability windows emerge for shorter lattice
sizes regime.}}

\section{Introduction}

Optical vortices represent unique solutions to the paraxial wave equation. These
beams demonstrate a distinct spatial layout, featuring an amplitude that reaches
zero at a single point within the plane perpendicular to the propagation axis.
At this specific point, the phase of the solution becomes indeterminate.  This
indetermination, or singularity, is encircled by a phase that twists around it,
accumulating a total phase equal to an integer number of $2\pi$, for a closed
circuit, as explained in reference~\cite{DESYATNIKOV2005291}. This integer
quantity is called the vorticity ($S$), or topological charge (TC), of the beam. In
recent times, vortex beams have garnered significant attention due to their
potential usefulness to encoding and storing information for applications in
optical communications~\cite{Barreiro:2008aa, PhysRevLett.88.013601,
Mair:2001aa}, as well as, in biophotonics, where they are useful due to their
capability to affect the movement and behavior of living cells, viruses, and
molecules~\cite{Zhuang188, Rodrigo:18, Favre-Bulle:2019aa}.  

Spatial soliton emerges as a self-trapped beam of light that retains its shape
while moving through a medium. The equilibrium between diffraction and nonlinear
effects enables the existence and propagation of these unique optical
structures~\cite{kivshar2003optical}.  These entities can be characterized as
self-localized solutions within nonlinear paraxial wave equations, predicted and
observed in diverse physical systems~\cite{RevModPhys.83.247}. Equations
commonly found in optics belonging to this category include the nonlinear
Schr\"{o}dinger Equation (NLSE), in which the linear potential models the
transversal distribution of the refractive index for the propagating media. 
These systems can be analyzed within the framework of the coupled-mode
approach. Nonlinear periodic structures offer alternative ways to control light
propagation by modification of diffraction properties through refractive index
modulation. The model that emerges from this analysis is referred to as the
discrete nonlinear Schr\"{o}dinger Equation
(DNLSE)~\cite{kevrekidis2009discrete}.  For conservative models described by the
DNLSE there is no allowance for the exchange of energy with the surroundings
which becomes reflected in a hermitic Hamiltonian, however Bender and his
collaborators~\cite{PhysRevLett.80.5243, PhysRevLett.89.270401} have
demonstrated that non-Hermitian Hamiltonians can still yield exclusively real
eigenvalues, provided the system exhibits invariance under combined operations
of parity $\mathcal{P}$ and time-reversal $\mathcal{T}$.  In optics, the
$\mathcal{PT}$-symmetry requires that real part of the
refractive index must be an even function, while the imaginary part must be an
odd function in space~\cite{SzameitPT2024, ruter-pt, Kartashov:16}. In a
$\mathcal{PT}$-symmetric system, the interplay of loss and gain can effectively
balance each other, resulting in a bounded dynamic behavior~\cite{Longhi:16,
PhysRevA.91.023822}.  Such a system can undergo spontaneous symmetry breaking,
transitioning from a $\mathcal{PT}$-symmetric phase with all real eigenvalues to
a broken phase with at least two complex eigenvalues, as the imaginary part of
the potential is increased~\cite{PhysRevA.91.033815}. 

Femtosecond laser inscription has emerged as a versatile technique for
fabricating optical waveguides in both amorphous and crystalline 
materials~\cite{Szameit_2010, Castillo:s}. This approach offers a unique
advantage: the ability to combine nonlinearity with strong variations in the
refractive index. Additionally, the same laser inscription technique can be
employed to selectively introduce losses within specific waveguides of the
array~\cite{Szameit2013}. Therefore, we believe discreteness, nonlinearity, and
$\mathcal{PT}$-symmetry represent a powerful combination of properties
achievable in experiments. Motivated by this potential scenario, this work explores the 
existence and stability of discrete vortex ring solitons within a 
$\mathcal{PT}$-symmetric DNLSE.

\section{Model}

The DNLSE, in its general form, can be written as
\begin{equation} 
    -i \frac{dU_{\vec{n}}}{dz} =\sum_{\vec{m}} C_{\vec{n},\vec{m}}U_{\vec{m}}
     + i\rho_{\vec{n}}U_{\vec{n}} + \gamma|U_{\vec{n}}|^{2}U_{\vec{n}},
\label{eq1} 
\end{equation}
where $U_{\vec{n}}$ is the electric filed amplitude at waveguide $\vec{n}$, $z$
is the coordinate along the longitudinal direction, $C_{\vec{n},\vec{m}}$ are
the coupling terms between sites $\vec{n}$ and $\vec{m}$. 
By restricting the interaction to nearest neighbors within the lattice and
setting the strength to the unity we simplify the coupling landscape to
($C_{\vec{n},\vec{m}} = 1$).
To analyze the effect
of energy exchange of the system with surroundings modeled in the $\mathcal{PT}$
symmetric manner, we introduce a linear potential with alternating gain and
loss. The strength and spatial structure of this potential becomes represented
by the $\rho_{\vec{n}}$ term. As we are going to deal with ring mode structure
and, in order to satisfy the $\mathcal{PT}$ symmetric requirements, we choose an
alternating behavior for the exchange of energy, i.\ e., the gain and losses in
our system will have an staggered form: $\rho_{\vec{n}} = (-\rho, \rho, ...,
-\rho, \rho )$, with $\rho$ being the gain/loss parameter. The lattice size
$L_s$ can be written as $L_s = 2N$, where $N$ is the total number of unitary
cells of the lattice.  The last term in Eq.~(\ref{eq1}) assumes the role of
nonlinear response of the system, that in our case corresponds with a cubic Kerr
type whose nonlinear coefficient is $\gamma$.

\section{Results}

We look for stationary solutions of model (\ref{eq1}) in the usual form,
$U_{\vec{n}}(z)=u_{\vec{n}} \exp (i \lambda z)$, where $u_{\vec{n}}$ is the
field amplitude which defines (in general) a complex spatial profile of the
solution, and $\lambda$ is the propagation constant. Under the coupled-mode
approach, optical ring vortex-type modes can be obtained by assuming periodic boundary
conditions in the one-dimensional version of Eq.~(\ref{eq1}). 

For the linear regime $(\gamma=0)$ the 
eigenvalue associated problem corresponds to

\begin{equation}
\lambda u_{n} = u_{n+1} + u_{n-1} + \rho_{n} u_{n},
\label{eq3}
\end{equation}

with $\lambda^{2}=4 V^2 \cos^2(\kappa/2)-\rho^{2}$ as the corresponding
dispersion relation, where, $\kappa=(2 \pi S / N)$, and $N$ is the total number
of unitary cells.  As we are modeling our circular array by imposing periodical
boundary conditions, our solutions also require that $u_{n}=u_{n+N}$. In order
to satisfy this condition $\kappa=(2 \pi / N) m$, being $m$ any integer number.
This also implies $m=S$.  Additionally, we can observe from dispersion relation
that when $\cos (\kappa/2)=0$, the propagation constant $\lambda$ becomes purely
imaginary, regardless of the value of $\rho$, signaling a completely unstable
mode. This condition is satisfied for $\kappa= q_{n} \pi$, with $q_{n}$ an odd
integer. For instance, for $L_{S}=8 \, (i.e., N=4)$ and $S=2$, the
quantized value of $\kappa$ is $\kappa=(2\pi / 4) \times 2$ which is equal to
$\pi $ for $q_{n}=1$.  This mode is unstable for any $\rho$ value. When $\kappa$
is not exactly equal to an odd multiple of $\pi$, there is a finite $\rho$ range
where all eigenvalues are real. This happen, for instance, for $L_{S}=8,
S=1$, where $\kappa=(2 \pi / 4) \times 1=\pi/2 \neq \pi  q_{n}$ for any odd
$q_{n}$. This mode is therefore, stable at low $\rho$ values. We see the
stability characteristics will exhibit oscillatory behavior with respect to the
lattice size and the topological charge of the modes.

For the nonlinear ($\gamma \neq 0$) regime we find the stationary modes
$u_{\vec{n}}(t)=u_{\vec{n}} \exp(i \lambda t)$ by solving the eigenvalue problem
\begin{equation}
    \lambda u_{n} = u_{n+1} + u_{n-1} + \rho_{n} u_{n} +
     \gamma \left|u_{n}\right|^2u_{n},
   \label{eq2}
\end{equation}
i.~e., a set of coupled nonlinear algebraic equations. Our initial calculations indicate that stable ring vortex solitons cannot be found in self-focusing media, regardless of the gain/loss parameter $\rho$.
Therefore, we assume a defocusing media by choosing a negative value for
nonlinear coefficient ($\gamma=-1$).  In order to find nonlinear annular vortex
beams we proceed to solve Eq.~(\ref{eq2}) by implementing numerically the Newton
iterative scheme. By choosing proper and reasonable seeds, in each case, we
unveil families of stationary solutions with different values of TCs. In
addition, we perform a standard linear stability analysis (see e.\,g.,
\cite{Maluckov2015}) on each solution reported in this
study. Instability is determined by the growth rate $g$, i.\ e., the maximum
imaginary part of the whole set of eigenvalues, therefore unstable solutions 
possess a $g > 0$ parameter. The stability of ring vortex structures in a
continuous conservative model, with rotating waveguides, was analyzed recently
by Kartashov~\cite{KARTASHOV2023113919}.

\begin{figure}[htb!]
    \includegraphics[width=0.8\columnwidth]{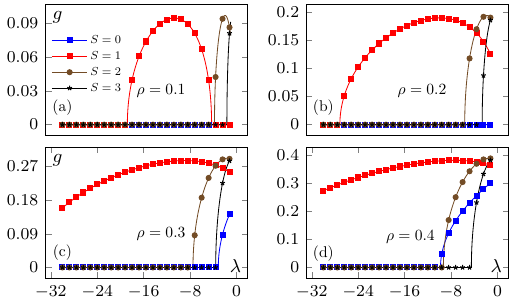} 
    \caption{$g$ vs $\lambda$ stability diagrams of ring modes with several
    values of vorticity (S) and gain/losses strength for a $L_s = 30$.  Blue
    line-squares for $S = 0$, red line-squares for $S = 1$, brown line-circles
    for $S = 2$ and black line-stars for $S = 3$, for (a) $\rho = 0.1$, (b)
    $\rho = 0.2$, (c) $\rho = 0.3$ and (d) $\rho = 0.4$.}
    \label{fig0} 
\end{figure}


Our analysis delves into the existence and stability characteristics of
vortex-type stationary modes within a nonlinear framework.
Figure~\ref{fig0}(a-d) portrays stability diagrams for distinct ring modes
classified by their $S$. Each diagram illustrates the relationship
between the instability gain ($g$) and the propagation constant ($\lambda$)
across diverse gain/loss regimes. A salient observation is the progressive
reduction in stability domains as the gain/loss strength escalates.
Surprisingly, the study also reveals a striking trend: modes possessing higher
$S$, characterized by more intricate swirling patterns, demonstrate
significantly larger stability domains compared to their simpler counterparts.
To analyze the vortex modes, we employed nonlinear plane waves as a baseline for
comparison. It is important to note that, formally, these plane waves correspond
to a ring vortex with zero vorticity ($S = 0$). For lower values of $\rho$,
these modes exhibit exceptional stability across the entire investigated
spectrum, remaining impervious to destabilizing forces.  For moderate and high
values of $\rho$ we observe that nonlinear plane wave destabilizes.
Figures~\ref{fig0}(c) and (d) demonstrate the instability of vortex ring beams
with topological charge $S = 1$. Interestingly, higher-order vorticity beams 
($S = 2$ and $S = 3$) exhibit stability under the same conditions. This behavior is
influenced by the lattice size ($L_s = 30$). As we will discuss later,
increasing vorticity can lead to the emergence of stable regions within shorter
waveguide arrays. We believe this is the underlying reason for the observed
stability of higher-order vortex ring modes.

In order to study the robustness for these ring vortex beams under a
$\mathcal{PT}$ symmetric energy exchange, we start by calculating stationary
vortex modes in absence of dissipation.  As we expected, the stability does not
depend on the length of the array.  However, in presence of alternating gain and
loss, the chain length becomes a critical parameter for stability. 
%
\begin{figure}[htb!]
    \includegraphics[width=0.8\columnwidth]{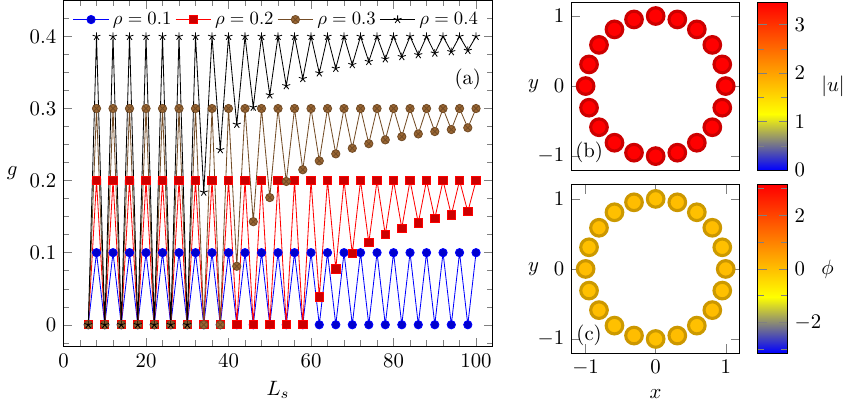} 
    \caption{(a) $g$ vs $L_s$ for ring modes with $S = 0$,
    at $\lambda = -10$. Blue line-circles for $\rho = 0.1$, red line-squares for
    $\rho = 0.2$, brown line-cross circles for $\rho = 0.3$ and black line-stars
    for $\rho = 0.4$.  (b) Amplitude and (c) phase profile for stable stationary
    solution corresponding with $L_s= 18$ and $\rho = 0.1$.}
    \label{fig1} 
\end{figure}
We begin our study analyzing the plane wave case ($S = 0$). We launch into our
Newton scheme a ring structure with constant amplitude. Figure~\ref{fig1}(a)
displays the behavior for the instability gain ($g$) in terms of length chain
($L$), for several values of gain/loss parameter. Similar to the linear
regime, instability in the nonlinear regime exhibits a periodic pattern with
respect to $L_s$. However, there exists a critical $L_s$ value beyond
which stable solutions disappear. Interestingly, this critical $L_s$ decreases as
the dissipation parameter ($\rho$) increases. In simpler terms, the regions of
stability shrink as the system experiences greater dissipation. This behavior is
evident in Fig.~\ref{fig1}(a), where the alternating pattern of stable and
unstable regions persists until a critical $L_s$ is reached.
Figure~\ref{fig1}(b) and (c) correspond with amplitude and phase profile for a
stationary stable solution with $L_s= 20$ and $\rho = 0.1$. 

\begin{figure}[htb!]
    \includegraphics[width=0.8\columnwidth]{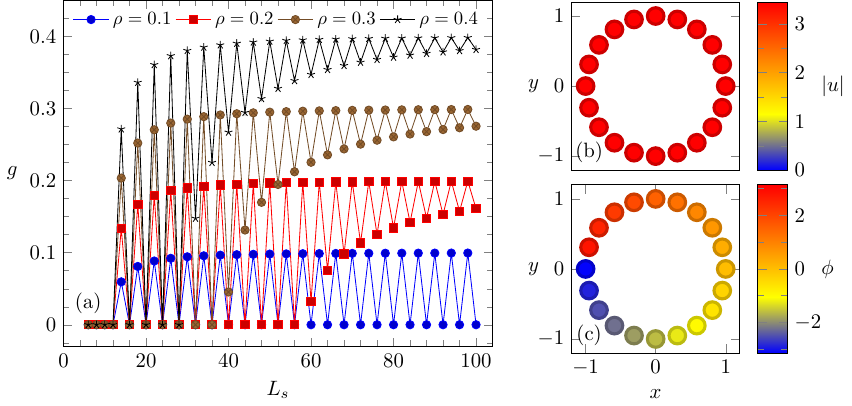}
        \caption{(a) $g$ vs $L_s$ for ring modes with $S = 1$,
        at $\lambda = -10$. Blue line-circles for $\rho = 0.1$, red line-squares for
        $\rho = 0.2$, brown line-cross circles for $\rho = 0.3$ and black line-stars
        for $\rho = 0.4$.  (b) Amplitude and (c) phase profile for stable stationary
        solution corresponding with $L_s= 20$ and $\rho = 0.1$.} \label{fig2}
\end{figure}

Continuing our analysis, we now investigate the stability of ring vortex modes
with $S = 1$. In this case, we utilize a ring structure with a constant
amplitude, endowed with a phase pattern varying from $-\pi$ to $\pi$, sampled
across the ring structure. Figure~\ref{fig2}(a) presents the behavior of the
instability gain ($g$) for the corresponding $S = 1$ scenario, expressed in
terms of chain length ($L_s$), across various values of the gain/loss parameter.
Since we are using a discrete model, there exists a minimum value for $L_s$ such
that the $2\pi$ quantity becomes adequately sampled; thus, our calculations
start at $L_s= 6$. An initial stage is observed where ring solutions up to $L_s
= 12$ consistently remain stable, regardless of the assigned value for $\rho$.
Subsequently, the behavior mirrors the $S = 0$ case, displaying a stability
oscillation up to a specific value for $L_s$, beyond which ring modes become
changeless unstable. Figure~\ref{fig2}(b) and (c) depict the amplitude and phase
profile, respectively, for a stationary, stable solution with $L_s= 20$ and $\rho
= 0.1$.

\begin{figure}[htb!]
    \includegraphics[width=0.8\columnwidth]{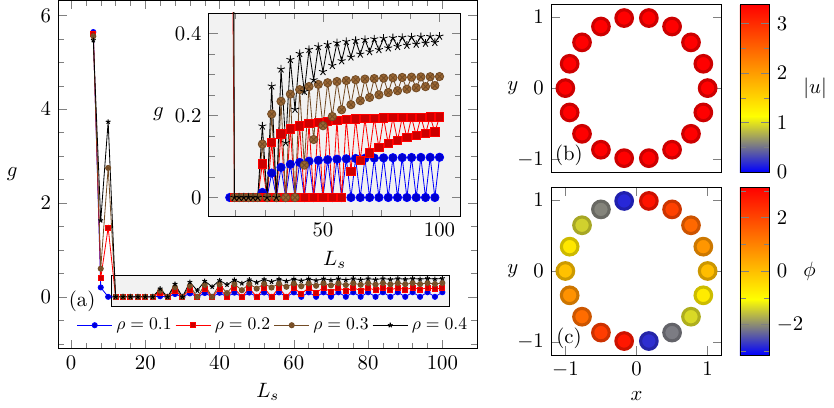}
    \caption{(a) $g$ vs $L_s$ for ring modes with $S = 2$,
    at $\lambda = -10$. Blue line-circles for $\rho = 0.1$, red
    line-squares for $\rho = 0.2$, brown line-cross circles for $\rho = 0.3$ and
    black line-stars for $\rho = 0.4$.  (b) Amplitude and (c) phase profile for
    stable stationary solution corresponding with $L_s= 18$ and $\rho = 0.1$.}
    \label{fig3}
\end{figure}

Now, our attention shifts to the stability of ring vortex modes with $S = 2$. In
this case, the initial configuration is equipped with a phase pattern varying
from $-2\pi$ to $2\pi$, spatially resolved across the ring. 
Figure~\ref{fig3}(a) depicts how the stability of this kind of mode ($S = 2$) is
affected by the interplay between the lattice size and the ($\rho$) parameter.
To guarantee sufficient sampling of the $4\pi$ phase shift around the ring, we
again chose a minimum $L_s = 6$.  Initially, we observe an unstable phase for
ring solutions up to a lattice size of $L_s = 12$. Interestingly, for $L_s$
values between 12 and 22, the solutions become consistently stable regardless of
the chosen dissipation parameter ($\rho$).  Following this trend, the behavior
for $L_s$ aligns with what we observed for $S = 0$ and $S = 1$ cases. Here, the
stability oscillates, with ring modes being alternately stable and unstable up
to a certain $L_s$ value.  Beyond that point, all ring modes become permanently
unstable.  Figure~\ref{fig3}(b) and (c) depict the amplitude and phase profile,
respectively, for a stationary, stable solution with $L_s= 20$ and $\rho = 0.1$.

\begin{figure}[htb!]
    \includegraphics[width=0.8\columnwidth]{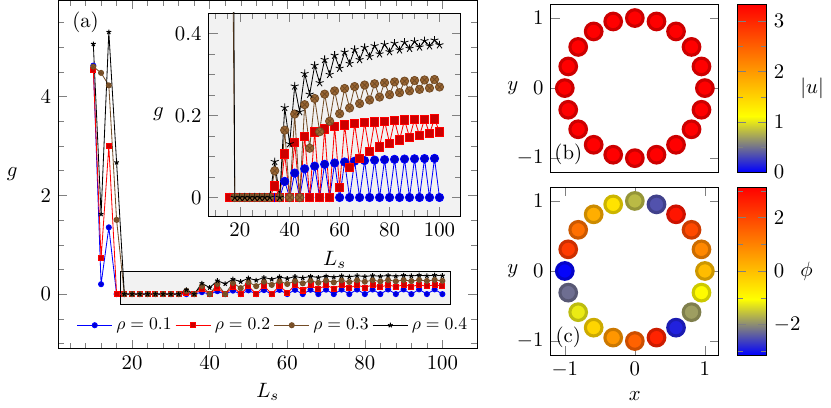}
    \caption{(a) $g$ vs $L_s$ for ring modes with $S = 3$,
    at $\lambda = -10$. Blue line-circles for $\rho = 0.1$, red
    line-squares for $\rho = 0.2$, brown line-cross circles for $\rho = 0.3$ and
    black line-stars for $\rho = 0.4$.  (b) Amplitude and (c) phase profile for
    stable stationary solution corresponding with $L_s= 20$ and $\rho = 0.1$.}
    \label{fig4}
\end{figure}

Stability in optical ring vortex modes with $S = 3$ exhibits a comparable
pattern. Here, the initial configuration features a phase profile that
continuously varies from $-3\pi$ to $3\pi$ across the ring. Figure~\ref{fig4}(a)
illustrates the instability gain ($g$) for the corresponding $S = 3$ scenario,
expressed in terms of chain length ($L_s$), considering various values of the
gain/loss parameter. As $S$ increases, it becomes imperative to augment the
minimum $L_s$ since sampling $6\pi$ demands more sites for vorticity to be well
defined.  Hence, we initiate with $L_s= 10$.  The analysis of ring soliton
stability reveals intriguing behavior. Initially, all modes exhibit instability
for lattice size up to $L_s = 18$. However, a window of stability emerges for
$L_s$ between 18 and 32. Within this window, all modes remain stable regardless
of the chosen dissipation parameter ($\rho$).  Beyond this critical $L_s$ value,
the behavior transitions back to what was observed for $S = 0$, 1, and 2 cases.
Here, stability oscillates with $L_s$, with modes alternating between stable and
unstable states until reaching a critical $L_s$.  After this critical point, all
ring modes become permanently unstable.  Figure~\ref{fig4}(b) and (c) depict the
amplitude and phase profile, respectively, for a stationary, stable solution
with $L_s= 20$ and $\rho = 0.1$.

\begin{figure}[htb!]
    \includegraphics[width=0.8\columnwidth]{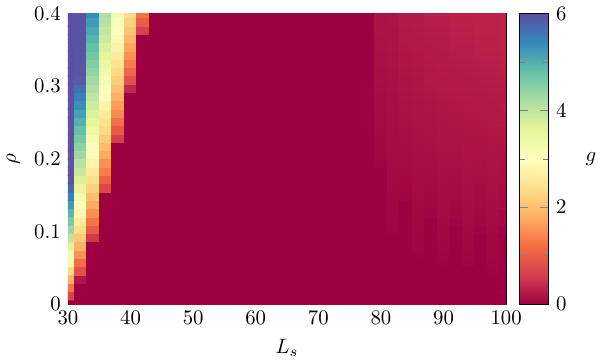}
    \caption{(a) Stability colormap diagram for ring modes with seven 
    topological charges ($S = 7$) in terms of $L_s$ and $\rho$.}
    \label{fig5}
\end{figure}

Recognizing the limitations of presenting examples within a single work, our
focus shifts to validating the existence of ring solitons with a higher value of
\(S\), specifically \(S = 7\).  We estimate the minimal length required to
support ring vortex beams with \(S = 7\), determining that a minimum of \(L_s=
30\) sites is necessary to obtain this type of solution. Within the range \(L_s=
[44,\,78]\), we observe a domain where the modes remain stable, as long as the
$\rho$ stays within our maximum range. For lower values of $\rho$, 
the oscillating stability pattern resumes beyond $L_s = 78$. However, 
for intermediate and higher values of $\rho$, the ring vortex modes enter a 
permanently unstable phase. These results are illustrated
in Fig.~(\ref{fig5}) through a colormap diagram, where instability gain 
$g$ is represented in terms of \(L_s\) and \(\rho\).

\begin{figure}[htb!]
    \includegraphics[width=0.8\columnwidth]{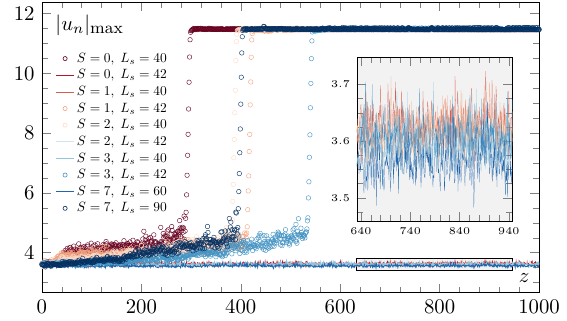}
    \caption{(a) Dynamics for maximum value of $|u_n|$ of 
    unstable (circles) and stable (lines) perturbed ring vortex modes.}
    \label{fig6}
\end{figure}
In order to corroborate predicted results displayed in 
Figs.~\ref{fig1}-\ref{fig5}, we perform a numerical integration of
Eq.~(\ref{eq1}) to propagate some of the optical ring vortex modes reported above. These
modes were perturbed by adding a small-amplitude white noise. By tracking the 
maximum value of the amplitude $|u_n|_{\textrm{\footnotesize max}}$, we  confirm
the stability behavior predicted for these ring vortex modes. Figure~\ref{fig6}
displays two cases, stable (lines) and unstable (circles), for modes with $S =
0,\, 1,\, 2,\, 3$ and 7 topological charges.  Initially, the maximum value of
unstable modes $|u_n|_{\textrm{\footnotesize max}}$ exhibits slow growth.
However, at a critical point, this value undergoes a dramatic increase. This
behavior reflects the interplay between gain and loss in the system. While the
optical power remains seemingly bounded for a while, it eventually reaches a new
steady state where gain and loss are balanced.


\section{Conclusions}

Summarizing, we have analyzed how the existence and stability become
influenced while an interplay of vorticity, lattice size and gain and loss, for
this kind of nonlinear waves.  Interestingly, for a fixed lattice size, our
findings suggest that vorticity counteracts the destabilizing influence of gain
and loss on these ring beams.  Furthermore, our predictions reveal that the
presence of gain and loss introduces an oscillating pattern of stability and
instability with respect to lattice size. However, instability dominates beyond
a certain threshold lattice size. Finally, our results demonstrate that
vorticity creates favorable conditions for the existence of specific ranges of
lattice size where stable ring vortex solitons are always stable.
This study elucidates the intricate interplay between vorticity,
gain/loss strength, and stability within a nonlinear system.

\section{Funding}
This work was supported by Fondecyt Grant No. 1200120.

\section{Acknowledgments}
This research was partially supported by
the supercomputing infrastructure of the NLHPC (ECM-02).


\section{Author Competing Interests}
The authors declare no conflicts of interest. 

\renewcommand\refname{References}

\end{doublespacing}

\end{document}